\documentclass[dvips]{acta} 
\usepackage{supertabular,lscape,epsfig} 
\usepackage{amssymb} 
\usepackage{amsmath}

\SetPages{0}{0}

\SetVol{1}{1}

\begin{document}

\begin{Titlepage}
\Title{The Optical Gravitational Lensing Experiment.\\
The OGLE-III Catalog of Variable Stars.\\
XII. Eclipsing Binary Stars in the Large Magellanic Cloud.}

\Author{D.~~G~r~a~c~z~y~k$^1$,\hspace{2pt} I.~~S~o~s~z~y~\'{n}~s~k~i$^2$,\hspace{2pt} R.~~P~o~l~e~s~k~i$^2$, \hspace{2pt}G.~~P~i~e~t~r~z~y~\'{n}~s~k~i$^{1,2}$,\\
A.~~U~d~a~l~s~k~i$^2$,\hspace{2pt} M.~K.~~S~z~y~m~a~\'{n}~s~k~i $^2$,\hspace{2pt} M.~~K~u~b~i~a~k$^2$,\hspace{2pt}
\L.~~W~y~r~z~y~k~o~w~s~k~i$^{2,3}$\hspace{2pt} and \hspace{2pt}K.~~U~l~a~c~z~y~k$^2$}
{$^1$Departamento de Astronomia, Universidad de Concepci\'{o}n, Casilla 160-C, Chile\\
e-mail:darek@astro-udec.cl\\
$^2$Warsaw University Observatory, Al. Ujazdowskie 4, 00-478 Warszawa, Poland\\
e-mail:(soszynsk,rpoleski,pietrzyn,udalski,msz,mk,kulaczyk)@astrouw.edu.pl\\
$^3$Institute of Astronomy, University of Cambridge, Madingley Road, Cambridge CB3 0HA, UK\\
e-mail:wyrzykow@ast.cam.ac.uk}

\Received{Month Day, Year}
\end{Titlepage}

\Abstract{We present catalog of 26\,121 visually inspected eclipsing
binary stars identified in the Large Magellanic Cloud during the third
phase of the Optical Gravitational Lensing Experiment. The sample is
limited to the out-of-eclipse brightness $I<20$~mag. The catalog
consists mostly of detached eclipsing binaries -- ellipsoidal variables
were not included. 

For stars brighter than $I=18$~mag the detection rate of eclipsing
binaries is 0.5\%  and for all stars it falls to 0.2\%. The absolute
completeness of the whole catalog is about 15\% assuming the occurence
rate of EBs toward the LMC equal to 1.5\%. 

Among thousands of regular eclipsing systems we distinguished a subclass
of eclipsing binaries -- transient eclipsing binaries (TEB) --
presenting cycles of appearance and disappearance of eclipses due to the
precession of their orbits.} 
{Methods: statistical, Stars: binaries: eclipsing, Stars: variables:
general}

\section{Introduction}
This is a successive paper presenting the variable stars treasury from
the third part of the Optical Gravitational Lensing Experiment (OGLE).
We focused here on the difficult task of identification of eclipsing
binary stars. In the past seven catalogs of eclipsing binaries (EBs) in
the Large Magellanic Cloud (LMC) found by the  microlensing surveys have
been presented: Grison \etal~(1995) found 79 candidate EBs from the EROS
survey, the MACHO survey identified 611 EBs (Alcock \etal~1997).
Derekas \etal~(2006) presented ``clean'' list of $3\;031$ EBs from MACHO
database. One year later Faccioli \etal~(2007) published an extension
of the preliminary catalog by Alcock \etal (1997)~containing $4\;634$
stars.

On the other hand, Wyrzykowski \etal~(2003) found $2\;580$ EBs in the
OGLE-II survey data. Additionally Groenewegen~(2005) and Graczyk \&
Eyer~(2010) identified 178 and 574 new EBs  respectively. In total,
using different approaches, $3\;332$ EBs were identified in the OGLE-II
photometric database. However, one should remember that this survey was
constrained mostly to the LMC bar. 

The OGLE-III survey covers a much larger area so we would expect larger
number of detected EBs. OGLE-II survey contained about 7 million sources
in the direction to the LMC (Udalski \etal~2000) while OGLE-III detected
about 32 million LMC sources. Using simple scaling we would expect to
find $\sim15\,000$ EBs. However, our catalog actually contains almost a
factor of two more objects. We discuss this result in Section 5 of our
paper.

\section{Observations and Data Reduction}

All the data presented in this paper were collected with the 1.3-m
Warsaw telescope at Las Campanas Observatory in Chile. The observatory
is operated by the Carnegie Institution for Science. During the
OGLE-III phase, the telescope was equipped with a mosaic eight-chip
camera, with the field of view of about $35\arcm\times35\arcm$ and the
scale of 0\zdot\arcs26~pixel$^{-1}$. For details of the instrumentation
setup we refer the reader to Udalski (2003).

116 OGLE-III fields in the LMC cover nearly 40 square degrees and about
32~million stars were detected on the collected images. Approximately
500 photometric points per star were secured over a timespan of eight
years, between July 2001 and May 2009. About 90\% observations were
taken in the standard {\it I} photometric band, while the remaining
measurements were taken in the {\it V}-band. The OGLE data reduction
pipeline is based on the Difference Image Analysis technique (Alard and
Lupton 1998, Wo\'{z}niak 2000, Udalski 2003). A full description of the
reduction techniques, photometric calibration and astrometric
transformations can be found in Udalski \etal (2008).

\section{Method of Identification}

Search for eclipsing binaries was done using the method outlined by
Graczyk \& Eyer~(2010). However, some changes to the method were
introduced. For $3\;332$ EBs detected in the LMC during OGLE-II survey
only one EB is fainter than $I\approx 20$~mag. The detection rate
falls very quickly for stars fainter than $I\approx 19$~mag.
Therefore, we limited our search of the candidate eclipsing binaries for
stars brighter than $I=20$~mag. Futhermore, all stars having less
than 120 measurements in the {\it I}-band were excluded from the search.
Thus, from a total number of 32 millions sources detected in the LMC,
only 12 millions sources were investigated for eclipsing-binary-like
variability. To save computational time the period search was restricted
to periods longer than 1.0015 day and shorter than 475 days with
$70\;000$ trial periods. Period searches were performed with the PDM
method (Stellingwerf~1978) for stars having a skewness parameter less
than 1.575, and with the string-length method (Lafler \& Kinman~1965)
for the remaining stars. Stars having  period longer than 6.45 days were
additionally investigated using the string-length method for periods
within the range $5.05-2800$ days, the last number being the
approximate time duration of OGLE-III project. 

\begin{figure}[htb]
\includegraphics[scale=0.7]{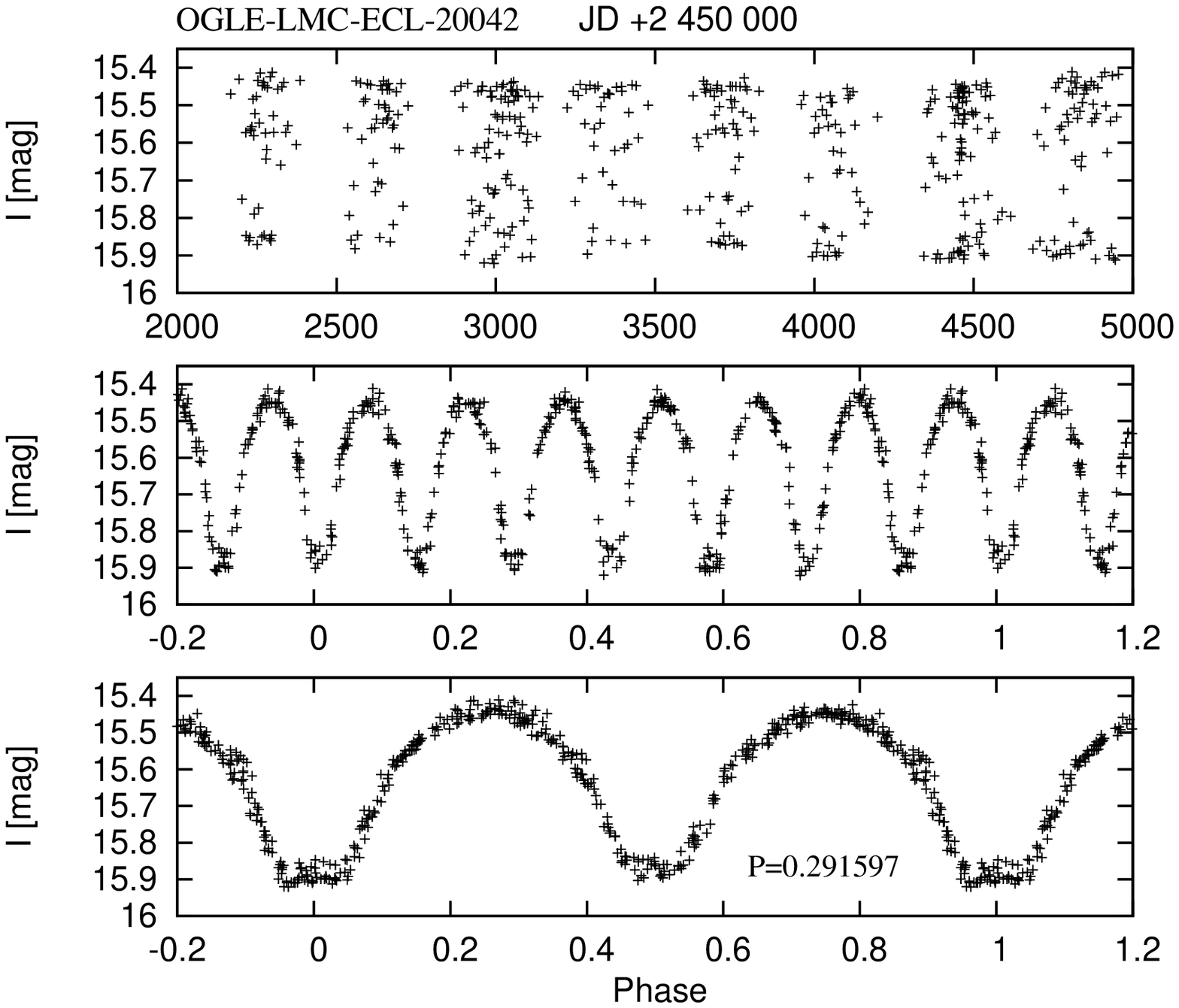}
\FigCap{Upper panel: light curve of OGLE-LMC-ECL-20042. Middle panel:
light curve of OGLE-LMC-ECL-20042 folded with the original period
1.02059 days found by our period finding algorithm. The star was
classified as an eclipsing binary candidate. During the visual
inspection it turned out that its true period is just 2/7th of the
original one. Bottom panel: light curve of OGLE-LMC-ECL-20042 folded
with the real period of 0.291597 days indicating a W~UMa type eclipsing
binary.}
\label{fig:1}
\end{figure}

From the beginning of our work, we were aware that a final visual
inspection of candidates would be necessary to produce a ``clean''
catalog of eclipsing binaries. For this reason we decided to look for
periods longer than 1.0015 days only. If an eclipsing binary has shorter
period we can still find it as a EB candidate. Such object is found with
a longer period being a multiplication of a real period -- see Fig.~1.
During the visual inspection most of these short period binaries were
easy to recognize and to assign the real period. In practice this method
worked reasonably well for EBs having orbital periods longer than 0.25
day because we have found only one eclipsing binary with a shorter
period. On the other hand we would expect a sharp cut-off in the
distribution for periods shorter than $\sim$0.2 days (Paczy\'{n}ski
\etal~2006). Inspection of Fig.~2 from Ficcioli \etal~(2007) can shed
some light on this. Their period distribution also has sharp cutoff near
the period of 0.25 days like our one (see Fig.~4 in this paper), however,
they could detect about 20 close eclipsing binaries with a period below
that limit. Thus our searching algorithm introduces a bias for the
shortest period binaries (periods $P<0.25$~day). However, because all of
them are foreground Galactic systems we can still consider our method
suitable for the LMC eclipsing binaries.

Quite large number of stars are semiregular long period variables
($P>500$~days) or high proper motion objects. When their light curve is
folded with trial periods, such stars often produce spurious detections
as eclipsing binaries, usually with periods being close to one day or a
multiplication of one day. We account for this effect by employing
strong filters for stars having periods close to a small whole number of
days: only stars with high variability index {\it pvi} could pass the
filter. Some real EB could have been removed from the sample this way
but much smaller sample of stars for visual inspection was a clear
adventage of this approach.              

All candidate stars were cross-checked against other previously
published catalogs of variable stars from OGLE-III survey, \ie Cepheids
(Soszy\'{n}ski \etal~2008a, 2008b), RR Lyr stars (Soszy\'{n}ski
\etal~2009a) and Long Period Variables (Soszy\'{n}ski \etal~2009b). The
purpose of this comparison was to remove from the candidate sample all
the remaining pulsating and long period semiregular variables. We
finished with a sample of about 79\,000 of candidate stars. 

The next and the most tedious part was a visual inspection. The
inspection was done using {\sc Vartool} program (written by
M.K.~Szyma{\'n}ski) which has a nice graphic interface. {\sc Vartool}
can show the raw and the folded light curve of candidate star at the
same time. The period used for folding light curve can be easily
modified, and each scrutinized star can be assigned a type of
variability. Verification of the orbital period was necessary for many
stars, usually by multiplying or dividing it by a factor of 2. Almost
14\% of candidate stars turned out to be ellipsoidal variables or their
artifacts, 35\% of the sample were false detections caused by some noise
in the photometry of fainter stars. 15\% turned out to be non-eclipsing
very long period variables and artifacts of pulsating stars or eclipsing
binaries. About $29\;000$ objects passed visual inspection as eclipsing
binaries. 

During visual inspection only the most clear artifacts of EBs were
removed. Some of the bright eclipsing binaries had as much as 5
artifacts which passed the inspection. To remove them, the
cross-correlation index of time series measurements was calculated for
all stars from a given OGLE-III field having similar period (within
0.2\%). All star pairs with such an index larger than 0.72 were
inspected visually to find which star is a true variable. The criteria
were: the quality of the light curve (more noisy light curves belong to
artifacts), the flux amplitude (the star having the larger amplitude is
probably the true variable) and the brightness (in about 90\% of cases
the true variable is the brighter one). Almost $1\;200$ objects turned
out to be artifacts of some neighboring EBs. Afterwards we checked all
the remaining stars from the sample against the presence of artifacts
using another method: those stars being in the same field within
2~arcsec of each other and having similar period (within 0.1\%) were
again inspected. In this manner we found about 500 more artifacts. 

The last part was cross-identification of eclipsing binaries from
neighboring fields. $1\;051$ stars were found in two fields and 16 were
found in three fields. The final number of $26\;121$ stars classified as
eclipsing binaries constitutes the present catalog.

\section{Classification and Basic Parameters} 

The catalog provides $26\;121$ entries, one for each detected eclipsing
binary star. For each EB we provide: 1) identification; 2) orbital
period; 3) epoch of the primary minimum; 4) mean out-of-eclipse {\it
I}-band magnitude; 5) $V-I$ color; 6) the depth of the primary minimum;
7) the standard deviation; 8) the skewness; 9) the kurtosis; 10) {\it
V}-band magnitude (from OGLE-III photometric maps), 11) {\it I}-band
magnitude (from OGLE-III photometric maps) 12) right ascension
(J2000.0); 13) declination (J2000.0); 14) the periodic variability index
{\it pvi}; 15) classification. Positions 7, 8 and 9 refer to the first
three statistical moments of the light curve.      

The orbital periods of all EBs from the catalog were refined using the
string-length method. Typical relative precision of the period
determination is about $\sim~10^{-4}$, but it varies considerably
depending on the light curve quality. For eclipsing binaries having only
one eclipse observed we provide the shortest possible orbital period.
The precision of the ephemeris varies from about 0.05\% to 2\% of the
orbital period. Mean out-of-eclipse {\it I}-band magnitudes were
calculated for orbital phases around 0.25 or/and 0.75. The $V-I$ colors
and coordinates were adopted from the LMC photometric maps. The
preliminary classification of EBs based on the Fourier series
coefficients of the light curves was done using LC\_CLASS program
written by Pojma\'{n}ski (2002). Only one fourth of the classified cases
were visually inspected, so a number of misclassified stars can be
expected. We traditionally divided EBs into three main subclasses
according to the light curve shape: detached (ED), semi-detached (ESD)
and contact (EC). Futhermore we distinguished some other types: ED/VAR -
detached with superimposed other kind of variability, ED/ESD -
detached/semidetached binaries, ED/TEB - detached Transient Eclipsing
Binaries (see Section~7), ELL/EC - ellipsoidal/contact binaries. The
last type was singled out by identifying eclipsing binaries having
ellipsoidal effects dominating their light curve and eclipses which are
usully very shallow and almost grazing. The catalog also contains 30
entries in common with the Double Periodic Variables catalog in the LMC
(Poleski \etal~2010). The eclipsing binaries with Cepheid components
which were detected previously by Soszy\'{n}ski \etal(2008a, 2008b) are
not included in the present catalog, \ie OGLE-LMC-CEP-0227, -CEP-1718,
-CEP-1812, -CEP-2532 with population I Cepheids and OGLE-LMC-T2CEP-021,
-T2CEP-023, -T2CEP-052, -T2CEP-077, -T2CEP-084, -T2CEP-093, -T2CEP-098
with type-II Cepheids.

\section{Detection Rate and Completeness of the Catalog}

\begin{figure}[htb]
\includegraphics[scale=0.7]{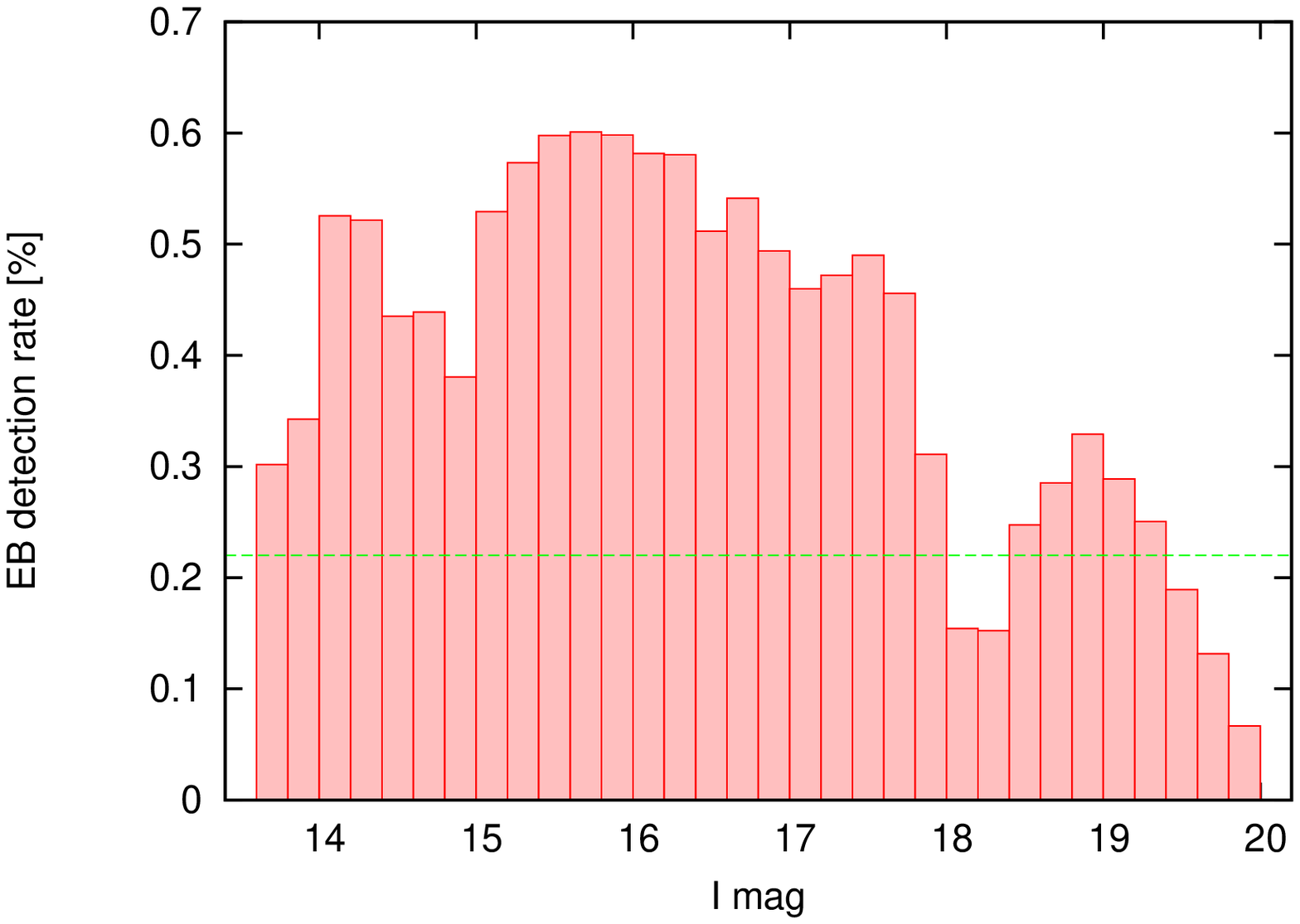}

\FigCap{Detection rate of eclipsing binaries in OGLE-III survey. The
mean value for the whole catalog is 0.22\% -- horizontal line. However,
for EBs brighter than $I\approx17.8$~mag the rate is substantially
larger and equal to 0.50\%. The dip around 18.2 mag is caused by
numerous red clump stars for which the detection rate is low. Also for
stars fainter than $I\approx 19$~mag one can note a clear deficiency of
the detection rate.}

\label{fig:2}
\end{figure}  

To assess the completeness of our catalog we employed two methods. The
first one providing ``absolute completeness'' is based on the estimated
total number of eclipsing binary systems brighter than our limiting
magnitude in all our observed LMC fields. To derive this number we
should know the occurence rate of eclipsing binaries in the LMC.
However, it is still an open question how double stars form and how such
processes depend on a population of stars, their metallicity and a
structure of particular galaxy. Therefore the occurence rate should be
rather estimated based on an empirical determination. 

Until now, the only reliable estimate of the ocurrence rate of eclipsing
systems was presented by Pr\v{s}a \etal~(2011) who calculated the
detection rate of EBs from the {\it Kepler} space mission to be about
1.5~\% in the direction of Cygnus-Lyra constellation. The population of
stars in the LMC is certainly different from that of the Galactic disk
(starbirth history, metallicity, space distribution), but both are
relatively young. If we assume the same rate of double stars, and
consistently EBs, in both populations and futhermore that the detection
rate from {\it Kepler} is a good estimation of the real occurence rate
we can make a first guess -- we would expect to find 180 000 EBs in LMC
from the OGLE-III survey. So the completeness of our catalog, in an
absolute sense, would be just 15\%. However, for stars brighter than
17.8 mag the number of detected sources is ~1.7 millions and the number
of identified EBs is about 8\,500 giving the detection rate of eclipsing
binaries of 0.50\% (see Fig.~2) and the absolute completeness of 35\%.
In fact, these ratios are even higher because a number of identified
sources, searched for EBs, are artifacts of bright stars. 

The second method providing ``relative completeness''  is based on the
number of EBs which could be detected from non-continuous, sparse
ground-based photometry. Our catalog was cross-correlated against the
MACHO catalogs of eclipsing binaries (Derekas \etal~2006, Faccioli
\etal~2007) and the OGLE-II catalogs of eclipsing binaries (Wyrzykowski
\etal~2003, Graczyk \& Eyer~2010). In the area covered by the OGLE-III
survey there are 2888 entries from Derekas \etal and 3819 entries from
Faccioli \etal~catalogs, respectively. It gives in total 4861 stars
because 1846 entries are common to both catalogs. 651 entries were not
found in our catalog. Inspection of the individual cases revealed that
most of them were ellipsoidal variables (excluded from our catalog by
definition) and only 205 were genuine EBs. That gives the relative
completeness of $\sim$ 95\%. Comparison with the OGLE-II catalogs shows
that 290 EB were not found in the present catalog which constitutes the
relative completeness of around 91\%. Taking a conservative limit, we
regard the completeness of our catalog at the level of 90\% in
comparison with previous ground based surveys.                     

There are $3\;332$ eclipsing binaries identified in the LMC from the
OGLE-II survey among $\sim$ 3.5 million sources brighter than $I\approx
20$~mag. This provides the detection rate of almost $\sim 0.1$\%. A
similar calculation for the OGLE-III survey in the LMC results in the
detection rate slightly larger than 0.2\%. What is the origin of this
difference?

First of all, the OGLE-III survey time span is twice that of the OGLE-II
one giving the chance to detect variability in larger number of stars,
especially those having longer periods. The number of measurements is
typically larger by a factor of only 1.25 per star in the OGLE-III
survey, but its photometry quality is superior to that of OGLE-II. This
again provides an opportunity to detect larger number of low amplitude
variables. Combining these two effects we can understand the higher
detection rate of the present catalog.

\section{Statistical Properties of the Catalog}

$16\;443$ entries from the catalog were classified as detached systems
(it constitutes almost 63\%), $1\;681$ entries as ED/ESD binaries (6\%),
$6\;502$ entries as semidetached systems (25\%) and only $1\;614$ as
contact or ellipsoidal/contact binaries (6\%). It is interesting to
compare these numbers with the numbers from {\it Kepler} Eclipsing
Binary Stars catalog (Pr\v{s}a \etal~2011, Stawson \etal~2011),
remembering that in our catalog ellipsoidal variables were excluded. If
we account for their missing contribution the distribution of the types
in the {\it Kepler} catalog is the following: 62\% detached EBs, 8\%
semidetached systems, 23\% contact binaries and 7\% of uncertain type.
It is worth noting that the relative number of detected, detached
systems is remarkably similar in both catalogs and very high in
comparison, and in clear contrast, with previous catalogs of eclipsing
binaries from ground-based surveys.

\begin{figure}[htb]
\includegraphics[scale=0.7]{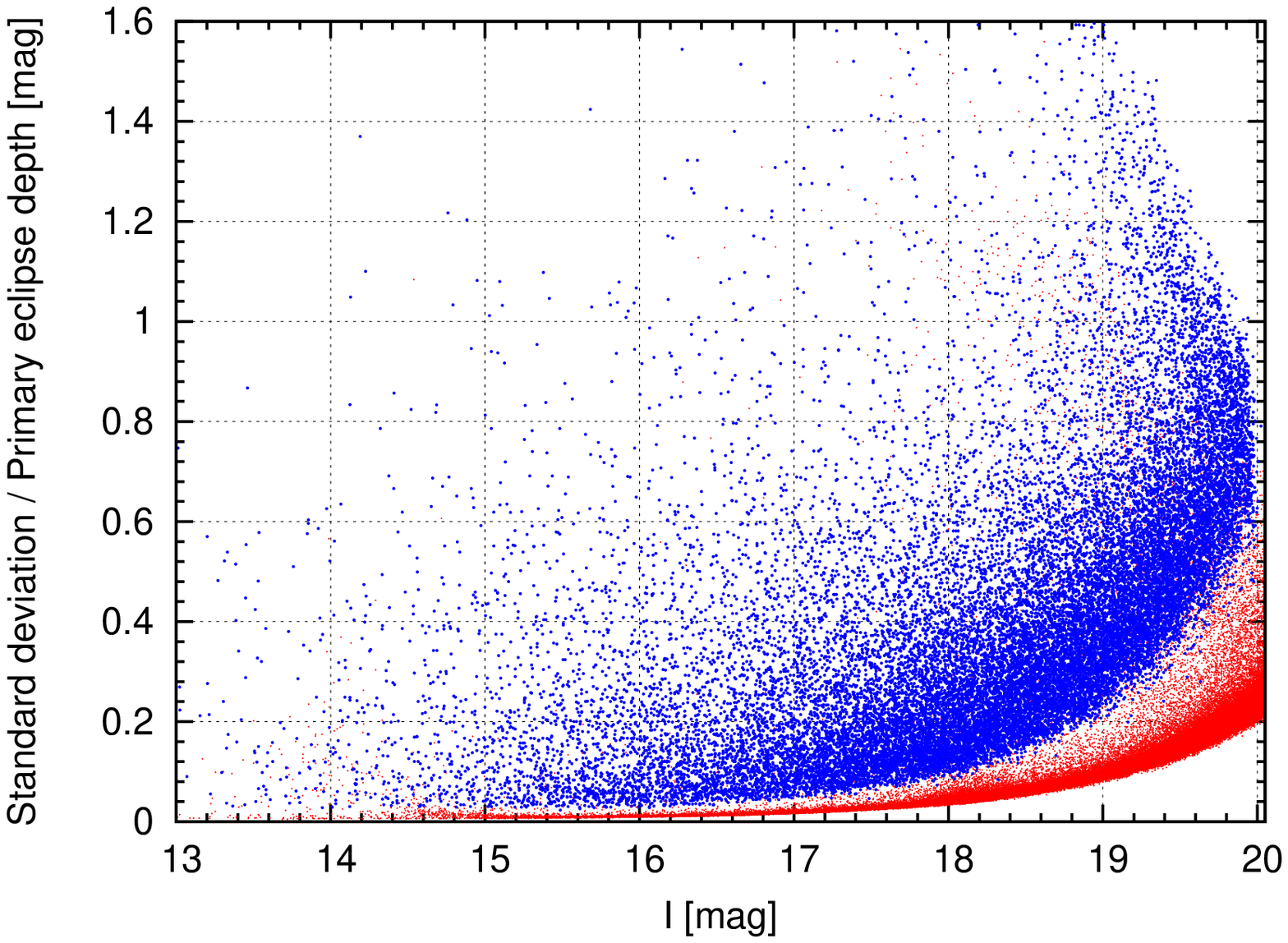}

\FigCap{Standard deviations ($\sigma$) \vs {\it I}-band magnitude for
all stars from LMC100.1 field (red dots) with superimposed primary
eclipse depths from our catalog (blue dots). We can interpret the bottom
of the $\sigma$ distribution as a noise limit for a given magnitude. On
the average we could detect eclipses having depth 2.5 times larger than
the noise limit.}

\label{fig:3}
\end{figure}

We believe that the conclusions given in Pr\v{s}a \etal~(2011) paper on
the low detection rate of detached binaries in previous catalogs 
\textquotedblleft stresses selection effects of ground based
surveys\textquotedblright~and that \textquotedblleft{\it Kepler's}
sensitivity to detached binaries is superior because of the continuous
data coverage\textquotedblright~are somewhat controversial. In our
opinion, sensitivity for detached binaries is conditioned mostly by the
use of the proper finding algorithm as we demonstrate in this paper and
if the time baseline and the number of measurements of the ground based
survey are large enough the selection effects are of minor importance.
Selection effects of previous catalogs were caused mainly by the use of
Fourier based period finding methods (poorly suited for detached EBs in
the case of not uniformly spaced data) and/or the lack of the efficient
filtering out of non-eclipsing stars at early stages of the catalog
preparation. We use such filtering based on the light curve statistical
moments which was proposed by Graczyk \& Eyer~(2010). The only real
selection effect here is that because of the observational noise
(caused, for example, by weather conditions) we are constrained to those
EBs which were showing relatively deep minima and lack strong additional
variability that smears out the presence of eclipses. For stars fainter
than $I=19$~mag the photometric noise in OGLE-III is larger than 0.1~mag
rendering detection of EBs with eclipses having depth smaller than
0.2~mag impossible (see Fig.~3). For the brightest stars we could detect
eclipses as shallow as $\sim$0.03 mag down to a limiting magnitude of
$I=16.5$.

There are, however, some differences between the distribution of EBs
from our catalog and the {\it Kepler} one. The relative number of
semidetached systems is much larger in our catalog. Part of this
disagreement may come from a different classification scheme. However,
we believe that there is an other reason behind it. In the LMC we probe
the upper part of the main sequence (O, B and early A type stars) where
numerous algol type semidetached systems exist. They can be easily
identified in our photometry because they have, usually, short orbital
periods and deep primary minima. On the contrary the relative number of
contact systems is much larger in the {\it Kepler} catalog. However,
most of them are short period main sequence binaries having absolute
luminosities well below the OGLE-III detection limit in the LMC.

\begin{figure}[htb]
\includegraphics[scale=0.7]{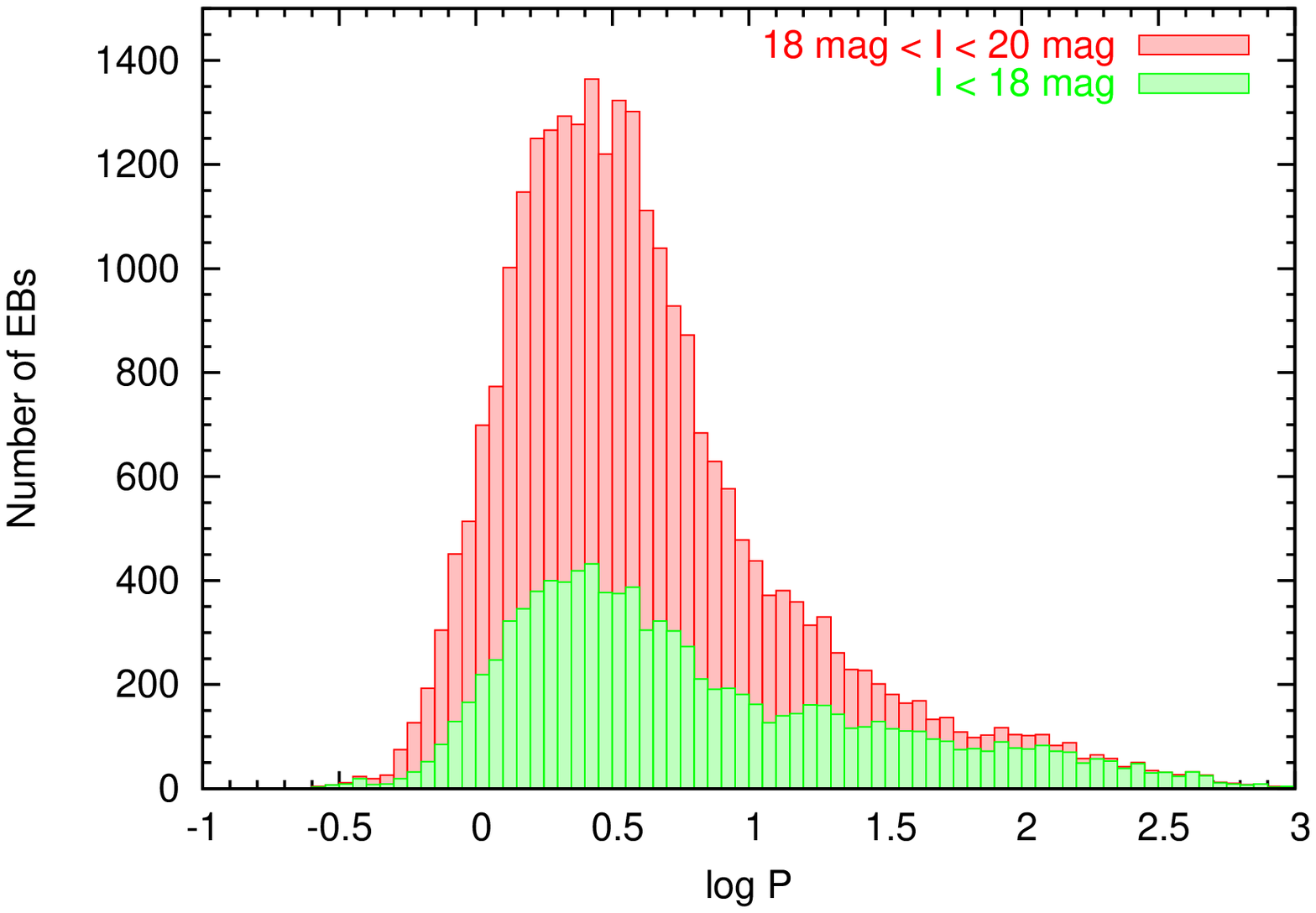}

\FigCap{Histogram of the EBs distribution as function of period
logarithm. It peaks for a period $\sim 2.5$~day. For stars brighter than
$I=18$~mag the distribution is much more flat what indicates that the
probability of finding an eclipsing binary of a given period depends on
its brightness. Detected eclipsing binaries fainter than 18~mag are, in
the vast majority, relatively short period binaries ($P<10$ days), and
at same time detected EBs with periods longer than 200 days are almost
exclusively systems brighter than 18~mag. Note the cut-off for systems
with orbital period shorter than 0.25 day and a power law decrease for
systems having period longer than 4 days with some excess visible around
100 days.}

\label{fig:4}
\end{figure}   

\begin{figure}[htb]
\includegraphics[scale=0.7]{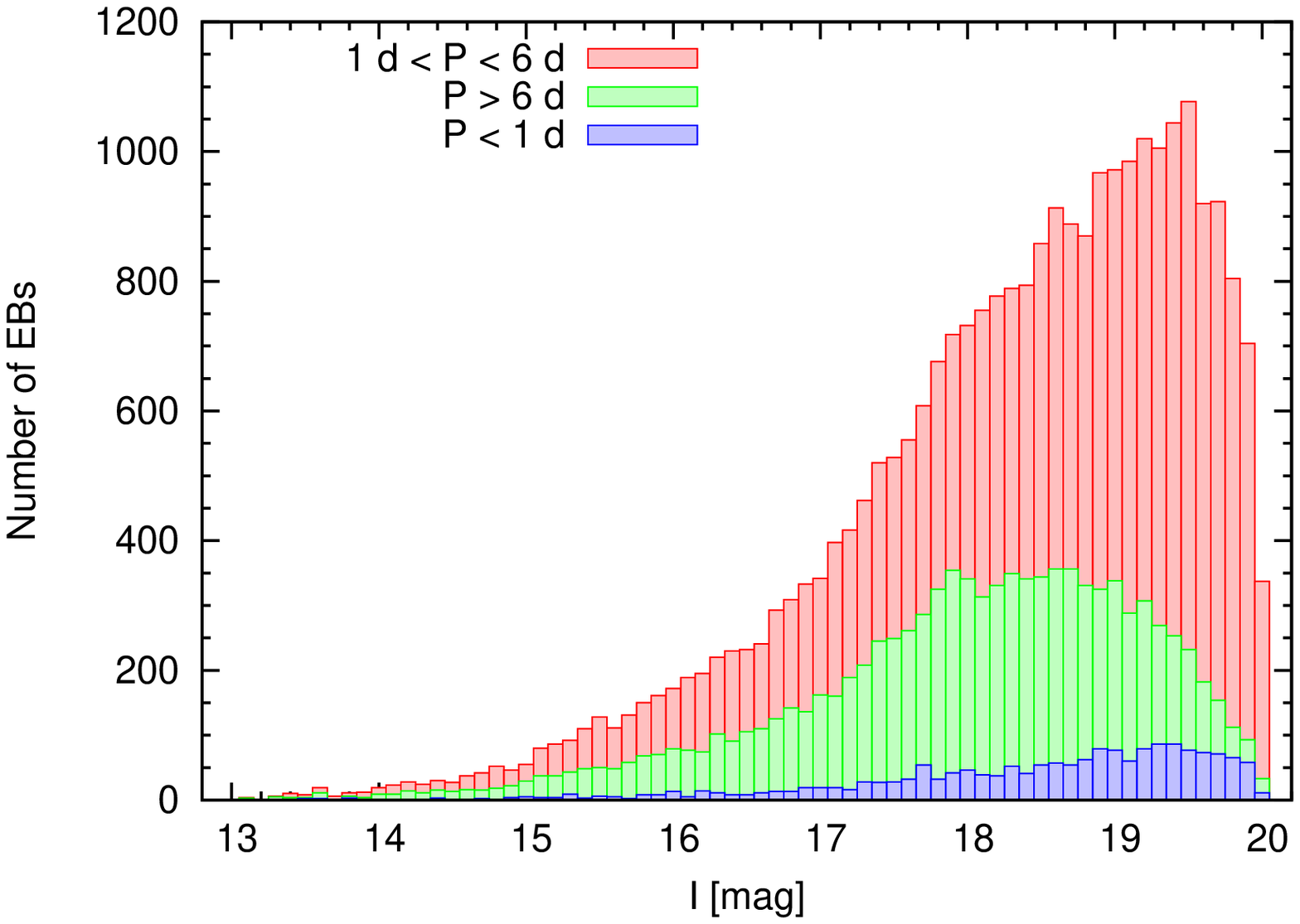}

\FigCap{Histogram of the number of detected EB in the {\it I}-band
magnitude bins. Note power law growth until around 18~mag and the peak
at $\approx 19.4$~mag again suggesting relatively good completeness of
the catalog for stars brighter than 18~mag. The distribution for very
short period systems ($P<1$~day) reflects that for all eclipsing
binaries. However, for longer period systems ($P>6$~days) the
distribution is different and after the power law rise it peaks already
at 17.8~mag.}

\label{fig:5}
\end{figure} 

Fig.~4 presents the distribution of orbital periods. There is one
distinctive peak of the distribution of all the stars from the catalog
at $\sim$2.5 days and it slightly differs from the MACHO LMC peak
distribution which is at $\sim$2.0 days (Derekas \etal~2007) and from
the ASAS Milky Way peak distribution ($\sim$1.5 days, Paczy\'{n}ski
\etal~2006). However, for stars brighter than $I=18$~mag the
distribution is somewhat more complex having additionally two smaller
peaks at around 15 days and 100 days.  

The number of detected EBs as a function of their brightness is
presented in Fig.~5. Systems brighter than 18~mag with periods of 6 days
or longer constitute 40\% of all stars. However, for fainter stars this
ratio is much lower: 25\% for $I=19$~mag and just 13\% for $I=19.5$~mag.
This clearly shows how the detection of long period EBs is biased for
faint stars. Fig.~6 shows the period-magnitude diagram for all detected
EBs. One can note the presence of narrow zone of ``avoidance'' close to
a period of 2 days. This is caused by the finding algorithm which has
strong filters to remove spurious candidates having periods of 1 day or
its multiples.

\begin{figure}[htb]
\includegraphics[scale=0.7]{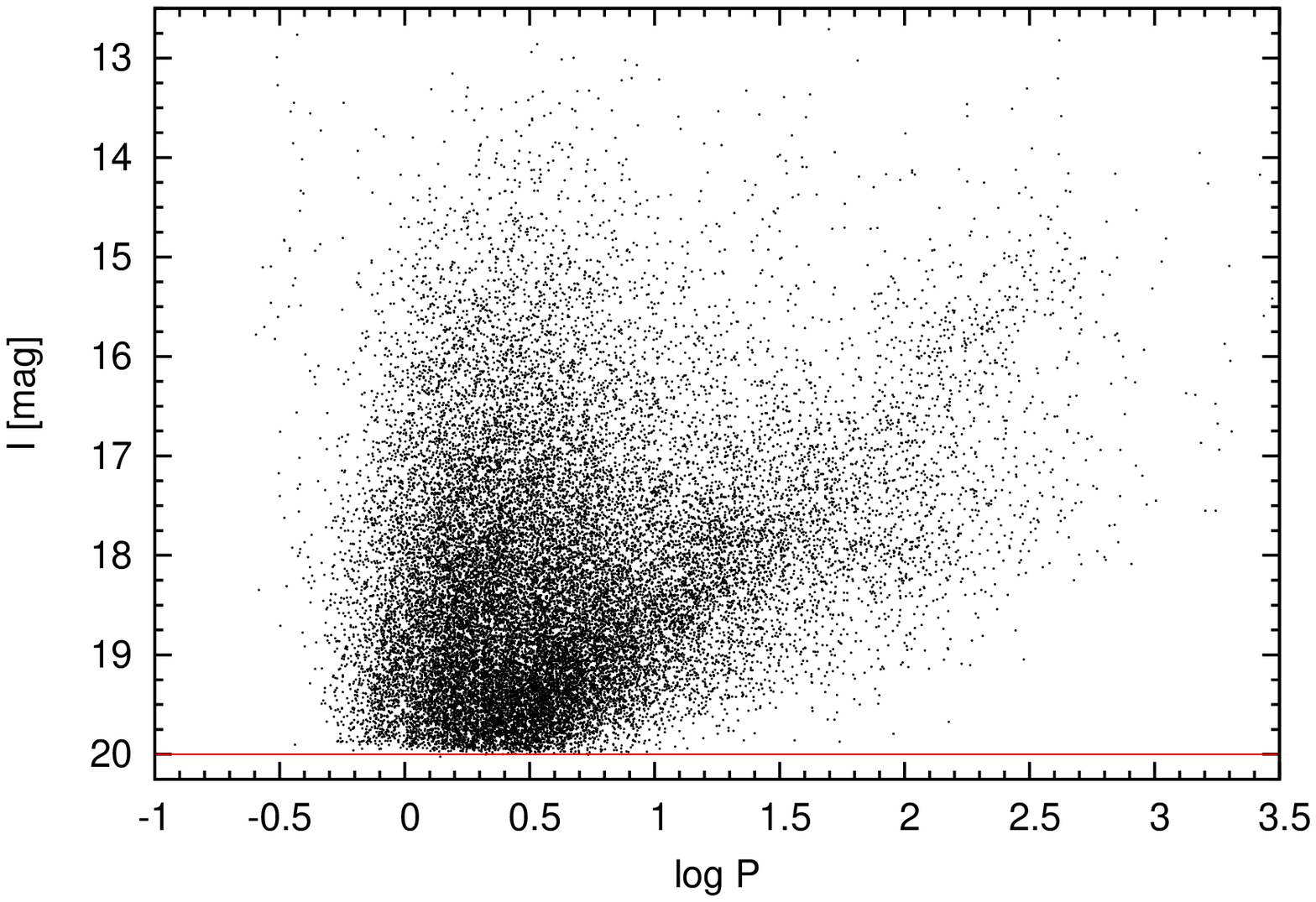}

\FigCap{{\it I}-band magnitude \vs logarithm of  period diagram for all
detected EBs. The vertical line at $I=20$~mag denotes the limiting
magnitude of the catalog.}

\label{fig:6}
\end{figure} 

\begin{figure}[htb]
\includegraphics[scale=0.7]{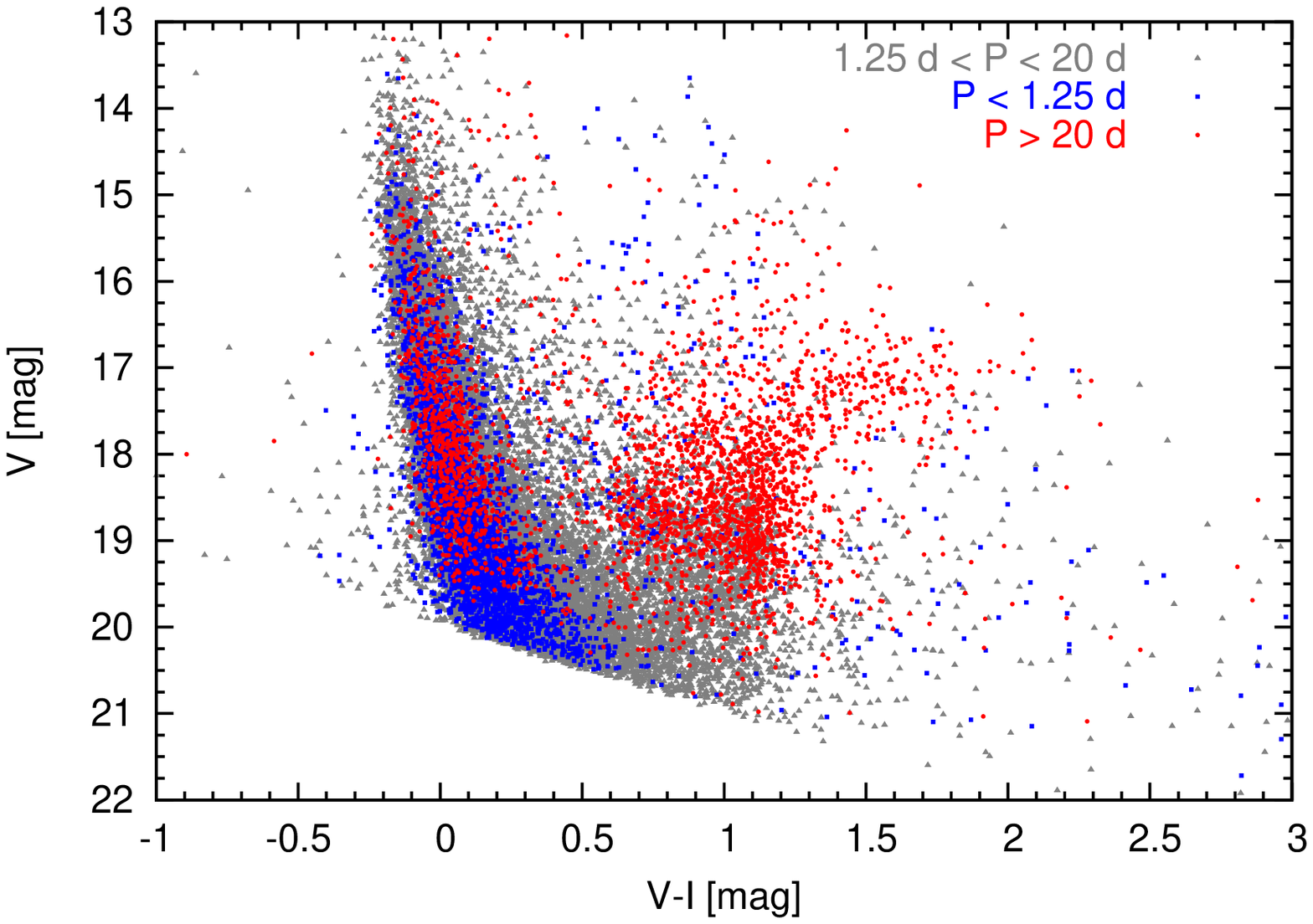}

\FigCap{Color-magnitude diagram for EBs from the catalog. Note the
bimodality of color distribution for systems with periods longer than
20~days: they clump around red giant branch and along the main
sequence.}

\label{fig:7}
\end{figure} 

The color magnitude diagram (CMD) is presented in Fig.~7. No
de-reddening was applied. Fig.~8 presents the color-period diagram. The
brightest stars clump in three separate regions: the LMC blue main
sequence, the LMC red giant branch and the Milky Way red, short period
close binaries. 

Fig.~8  also reveals a striking feature -- a lack of bright
($I<16$~mag), blue, long period EBs with periods $P\gtrsim$150 days. At
the first glance it seems that almost all long period progenitors
containing bright (M$_{V}<-2.0$) and massive ($M\gtrsim7.5 M_{\odot}$)
components have already evolved into red giant phase, but their shorter
period counterparts still remain on the main sequence. A question arises
if this is merely an observational bias as we could detect even fainter
($I>16$~mag), blue, long period EBs. 

The number of stars detected by OGLE-III in the LMC in the range of $16
< I < 18$~mag is six times larger than a number of the brighest stars
with $I<16$~mag. On the other hand,  counting blue, long period systems
in Fig.~8 we obtain this ratio equal to eleven, \ie almost two times
larger than if in the case of a simple selection effect. 

We can also count main sequence blue, bright systems (defined as
$V-I<0.2$) and, separately red giant, bright ones ($1.2< V-I < 1.6$) and
compare these numbers with stellar evolution expectations. We have 1 and
96 systems, respectively. This ratio for massive stars depends on
relative size between main sequence and red giant branch star, a
duration of these evolutionary phases, a mass loss rate shifting shorter
period sytems into $P\gtrsim150$~days region and a change in spectral
energy distribution during evolution shifting fainter (less massive)
stars into bright ($I<16$~mag) region during RGB phase. All above
mentioned factors are somehow uncertain (especially the mass loss rate)
but assuming reasonable limits we obtain the expected ratio between 1/50
and 1/20, again suggesting the lack of blue, long period systems. 

We propose two purely phenomenological explanations of this finding: 1)
the rate of stellar evolutionary processes in massive detached EBs
depends on the separation of binary components and systems which are
closer evolve in general slower; 2) the LMC long period systems, with a
large separation of their components, were, in general, formed before
their short period counterparts. However, we still cannot exclude an
existence of some hidden selection effects.

\begin{figure}[htb]
\includegraphics[scale=0.7]{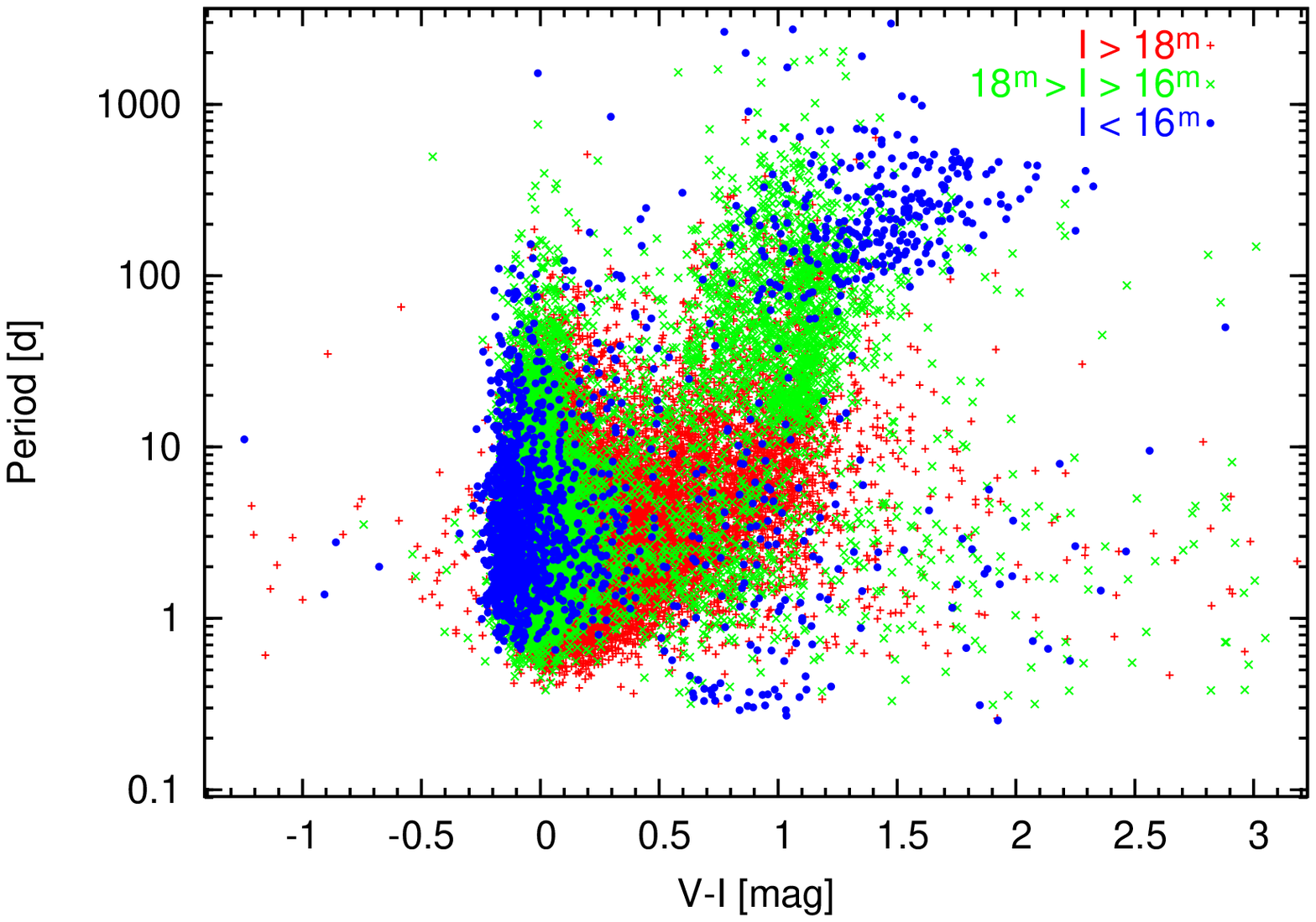}

\FigCap{Orbital period \vs $V-I$ color diagram. Note numerous blue EBs
with periods longer than 20~days. A comparison with the bottom panels of
Fig.~30 from Faccioli \etal(2007) demonstrates how the selection effects
are important in the detection of long period well detached binaries. }

\label{fig:8}
\end{figure}

\section{Interesting Eclipsing Binaries in the Catalog}  

While inspecting the EBs from the catalog we found 17 systems showing
the presence of a fast orbital precession. All of them are well detached
binaries and most of them are eccentric systems. They have blue colors
consistent with possessing late B-type or early A-type components. In
the most cases we see a gradual change of the depth of both eclipses: at
some moment of time the eclipses becoming visible, and then -- little by
little, deeper. Eventually, the eclipses become gradually shallower and
completely disappear. In two cases we observed the appearance of
eclipses and their disappearance during the time span of the OGLE-III
survey -- see Fig.~9. The apsidal movement as an explanation of such a
behavior can be ruled out because we observed two eclipses
simultaneously. We call such EBs 'Transient Eclipsing Binaries' (TEB).
They are virtually non-variable systems, but for some limited amount of
time we observe them as eclipsing binary stars when due the precession
of its orbital plane, or ``regression of the nodes'' (Soderhjelm 1975),
the inclination of their orbits to the line of the sight become close to
90 degrees. As a result we observe cycles during which the eclipses show
up and then gradually disappear for some period of time.

The prototypes of such TEBs in the Galaxy are SS Lac (discovery: Zakirov
\& Azimov 1990; analysis: Milone \etal~2000, Torres \& Stefanik 2000,
Torres 2001, Eggleton \& Kiseleva-Eggleton 2001) and V907 Sco
(discovery: Rahe \& Schoffel 1976; analysis: Lacy \etal~1999). The cause
of the orbital precession or changes of the orbital orientation is the
presence of a third, outer star being on inclined orbit to the inner
binary. The fast change of the eclipses depth in the LMC TEBs suggests that
the likely cause is ``regression of the nodes'' where the angular
momentum vector of the inner binary precesses around the total
(constant) angular momentum vector of the system. In the case of V907
Sco the nodal period of the cycle is just $\sim$70 years (Lacy
\etal~1999) and for most of the identified TEBs from the LMC it should
be of comparable length. It is worth noting that an eclipsing binary KID
5897826 found by the {\it Kepler} mission and having tertiary eclipses
also shows the presence of orbital precession caused by a third, outer
component (Carter \etal~2011).

\begin{figure}[htb]
\includegraphics[scale=0.7]{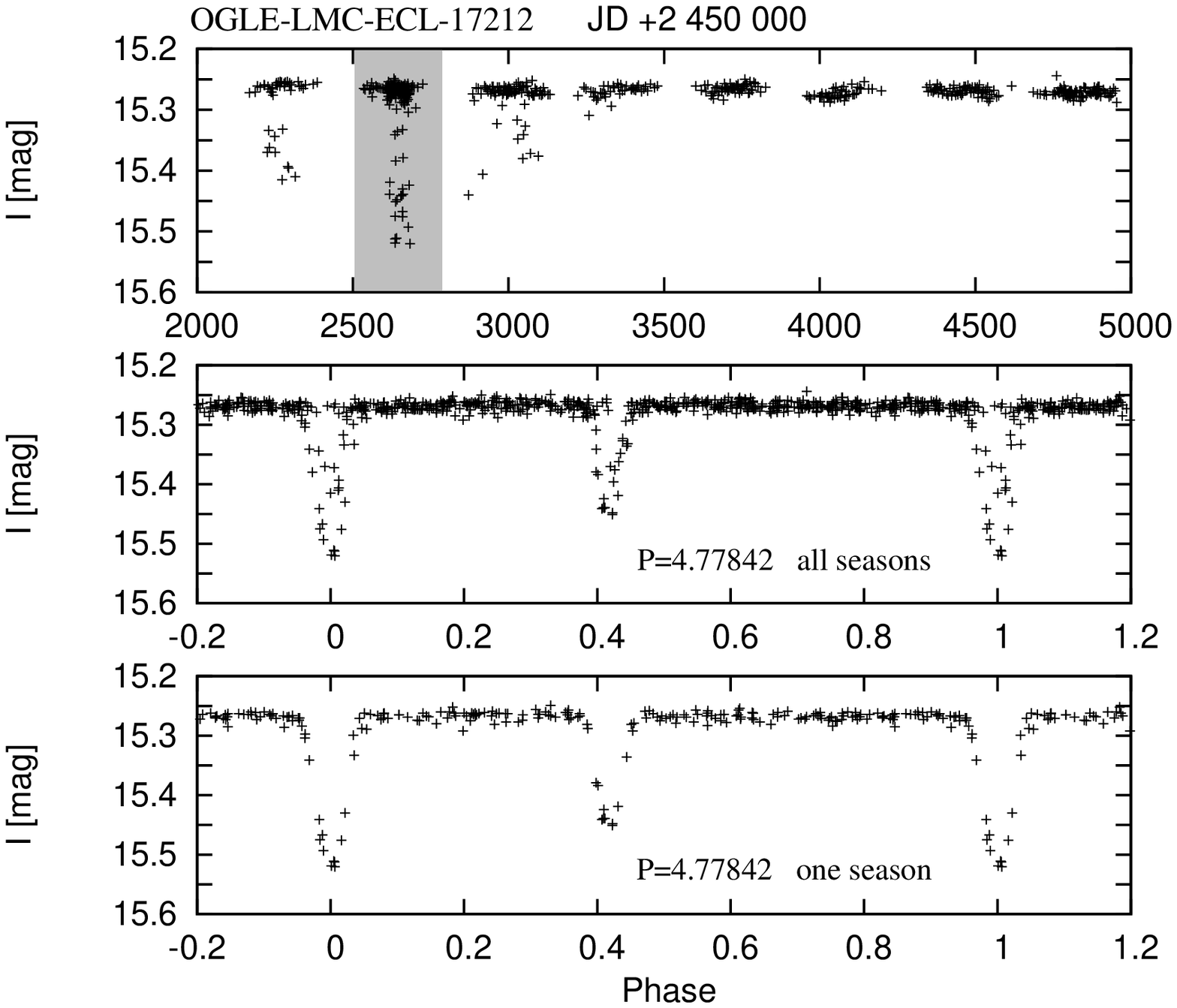}

\FigCap{Upper panel: an example of light curve of TEB system:
OGLE-LMC-ECL-17212. The time range between JD 2452500 and JD 2452800
when the eclipses were deepest is marked by a shadow. Note the lack of
eclipses after JD 2453400. Middle panel: light curve of
OGLE-LMC-ECL-17212 folded with the period P=4.77842 days. Bottom panel:
light curve folded with the same period but observations taken only from
the shadowed region in the upper panel; note the clear presence of well
defined eclipses in the eccentric system.}

\label{fig:9} 
\end{figure}  

Another group of interesting eclipsing systems are EBs having other kind
of variability superimposed and classified as ED/VAR. For example,
OGLE-LMC-ECL-02594 is EB containing (or being blended with) a bumper --
see Fig.~10. The astrometric position of the star during the brightening
is the same as during the baseline. Futhermore, the system lies in one
of a least crowded OGLE-III fields -- LMC131.7 -- strongly suggesting
that the bumper is one of the components of the binary and not only a
blend. 

OGLE-LMC-ECL-16549 is a long period eclipsing binary showing a
variability characteristic for a short period, contact eclipsing binary
(Fig.~11). Small dispersion of points in the primary minimum suggests
that the short period system is a blend or a part of a hierarchical,
gravitationally bounded system of at least four stars: two components
forming the long period system and two others in a contact system.
However, we do not detected astrometric shifts during deep eclipses.

OGLE-LMC-ECL-23999 is another long period system having large and
variable brightening visible after periastron passage (Fig.~12). Because
of the position of eclipses and their similar time duration we expected
that periastron passage was near the orbital phase 0.95, somewhere
between secondary and primary minimum. However, the brightening, which
could be interpreted as a result of intensive mutual reflection when
stars are close each other near periastron, occurs later after primary
minimum. Futhermore, the brightening is of considerably different
strength during consecutive passages. We suppose that they are related
to episodic mass exchange between components near periastron passage and
soon afterwards. Indeed, it seems that a sum of the radii of both
components is close to their smallest orbital separation.  

Fig.~13 presents observations of OGLE-LMC-ECL-17782 system. The light
curve of the system possess a number of odd looking features. First note
a strange, wide, flat-bottomed primary eclipse, reminiscent of
$\epsilon$ Aurigae eclipses, with additional narrow, eclipse-like
feature in the middle of the eclipse. There is also a considerable
change in the primary eclipse shape. Second, note the numerous downward
outliers which were not caused by worse weather conditions and we
consider them to be real. Third, around the phase 0.5 there is a narrow
secondary minimum, strikingly shorter than the primary minimum, but of a
similar duration as the eclipse-like feature visible near the orbital
phase 0. We interpret these features as the following: the circular,
detached or semidetached system contains two stars, one of them being
partially hidden within a semi-transparent, dark, elongated body or a
disk. When the disk transits over the second star, ``boxed'' shaped
minimum is produced and when the first star occults the second one, we
can see additional fading near the phase 0. When the second star
transits over the disk, we can detect only a narrow secondary eclipse
which corresponds to an occultation of the first star (the disk itself
does not contribute significantly to the total light, at least not in
the {\it I}-band). The disk has probably time-varying dimensions and
density producing different shapes of primary minimum during consecutive
seasons. Futhermore there are transient structures in the system (disk
debris?) responsible for additional minima at different orbital phases
when one of the stars is hidden behind them. Probably this structure is
somehow related to the non-uniform mass exchange between components. The
real puzzle is how such a dark, semitransparent disk-like structure
could be formed in such a short period system?   

OGLE-LMC-ECL-17681 is the shortest period system ever detected in the
direction of the LMC (Fig.~14). The orbital period is just 2 hours. The
system lies in the vicinity of the LMC bar, but its brightness
($V=19.2$~mag), color ($V-I=0.5$~mag), fast proper motion and very short period strongly
suggest that it is a Galactic object. Probably the system contains a white dwarf and a red dwarf.   

An interesting pair of short period, low mass EBs is presented in
Fig.~15. The systems are separated in the sky by only 15 arc minutes.
Both have a very red color: $V-I=2.88$~mag, and their light curves are
very similar. The relative difference of their periods is only 5$\cdot
10^{-4}$. However, the epochs of the primary minimum are different
indicating that they are not artifacts of the same eclipsing binary.
Most likely, they are foreground Milky Way objects comprising a wide,
hierarchical quadrupole system.

\begin{figure}[htb]
\includegraphics[scale=0.7]{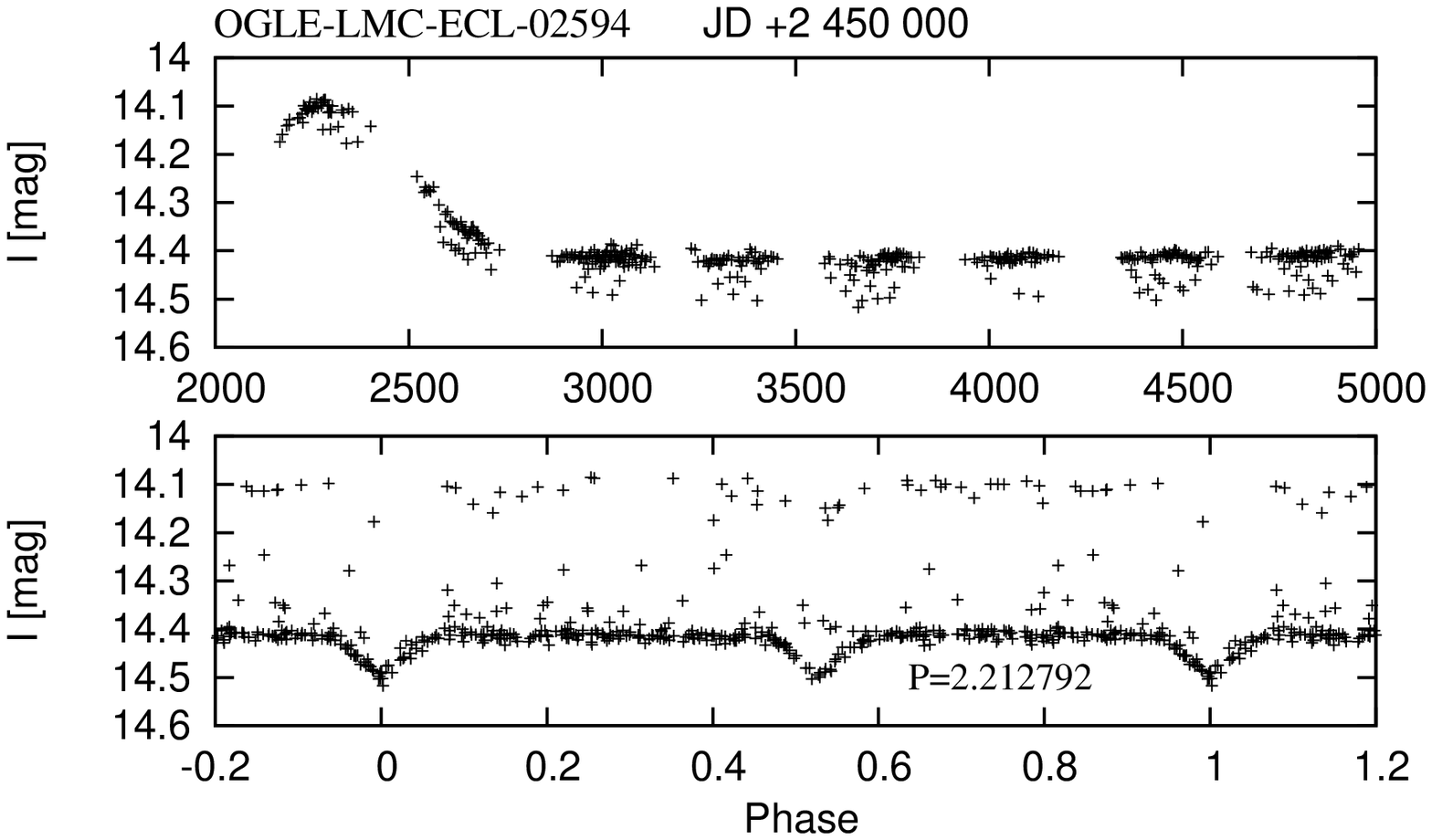}

\FigCap{Upper panel: light curve of OGLE-LMC-ECL-02594. Note the
brightening during the first two OGLE-III observational seasons. This
brightening is characteristic of a bumper variable. A blue $V-I$ color
of OGLE-LMC-ECL-02594 supports such interpretation. Bottom panel: light
curve of OGLE-LMC-ECL-02594 folded with the period P=2.21279 days
showing well defined eclipses. }

\label{fig:10} 
\end{figure}

\begin{figure}[htb]
\includegraphics[scale=0.7]{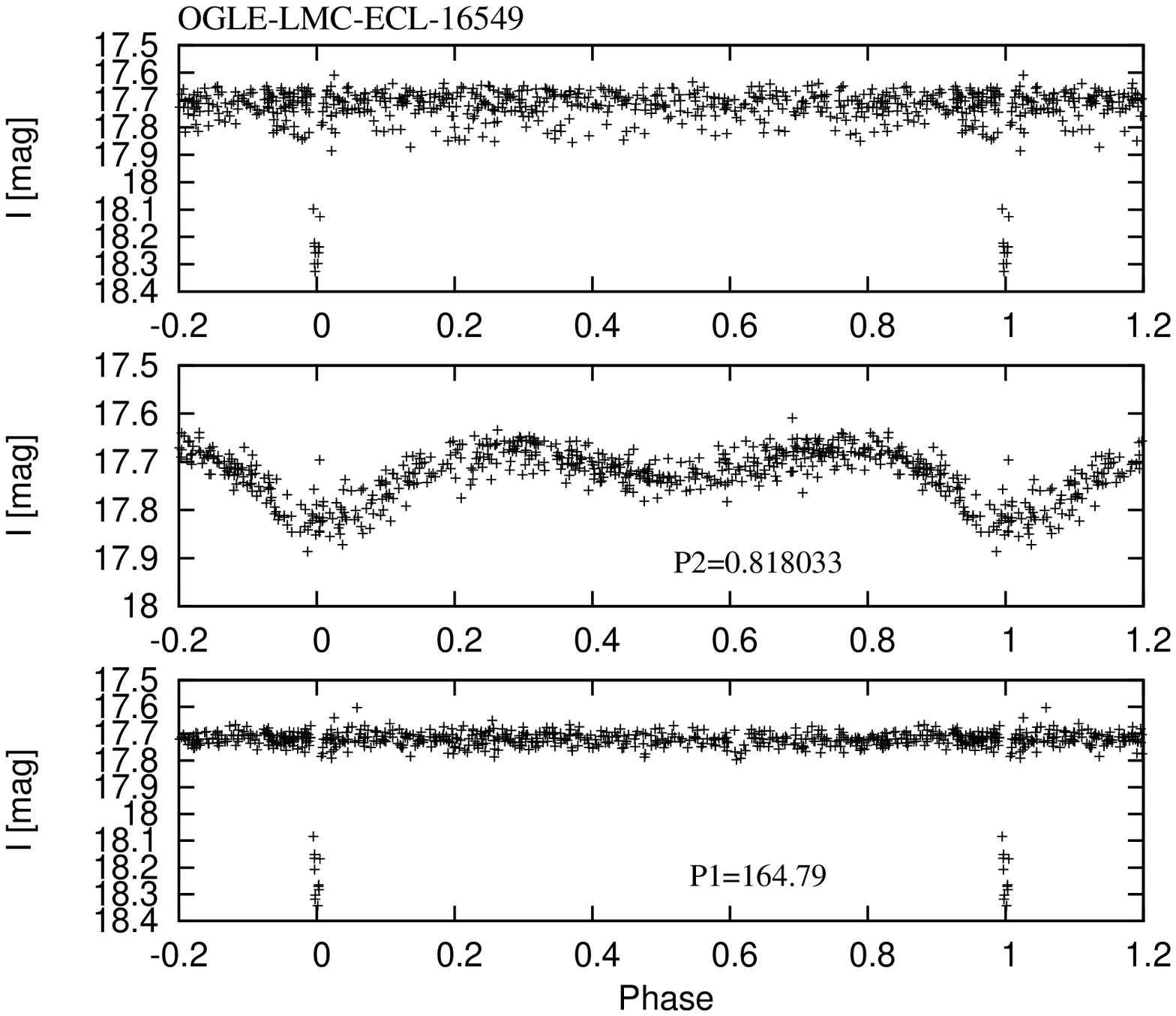}

\FigCap{Upper panel: light curve of OGLE-LMC-ECL-16549 folded with
the orbital period $P_1=164.79$~days. Note the numerous downward outliers
outside the deep, narrow primary eclipse. Middle panel: light curve of
OGLE-LMC-ECL-16549 filtered out of deep eclipses and folded with the
period $P_2=0.818033$~days. This reveales the presence of a contact binary.
Bottom panel: light curve of OGLE-LMC-ECL-16549 filtered out of short
period variability and folded with period $P_1$. Note the trace of a
possible secondary minimum near the orbital phase 0.6.}

\label{fig:11} 
\end{figure}

\begin{figure}[htb]
\includegraphics[scale=0.7]{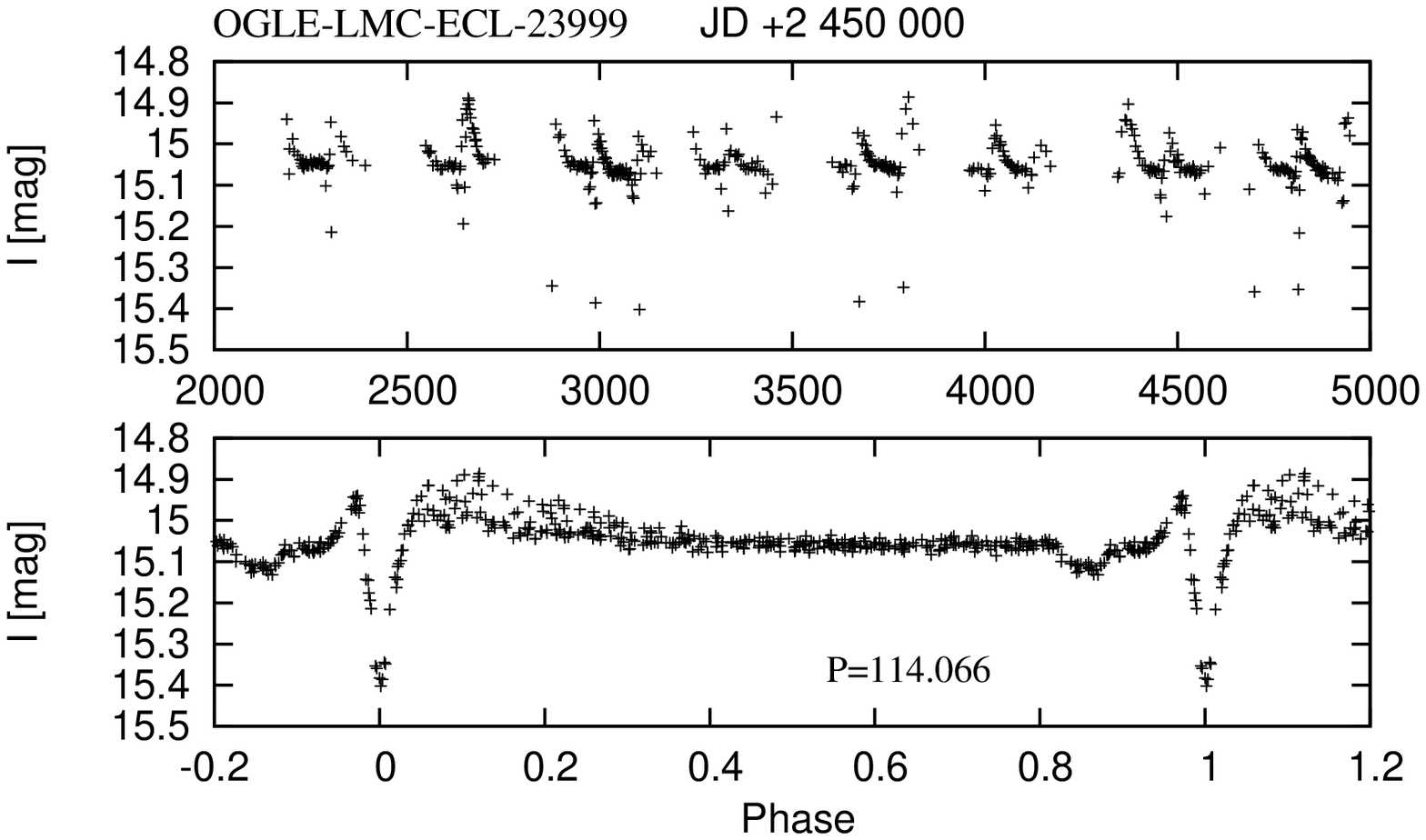}

\FigCap{Upper panel: light curve of a long period system
OGLE-LMC-ECL-23999. Bottom panel: light curve of OGLE-LMC-ECL-23999
folded with the period $P=114.066$~days showing an eccentric system with
relatively wide eclipses and fairly large proximity effects.}

\label{fig:12} 
\end{figure}

\begin{figure}[htb]
\includegraphics[scale=0.7]{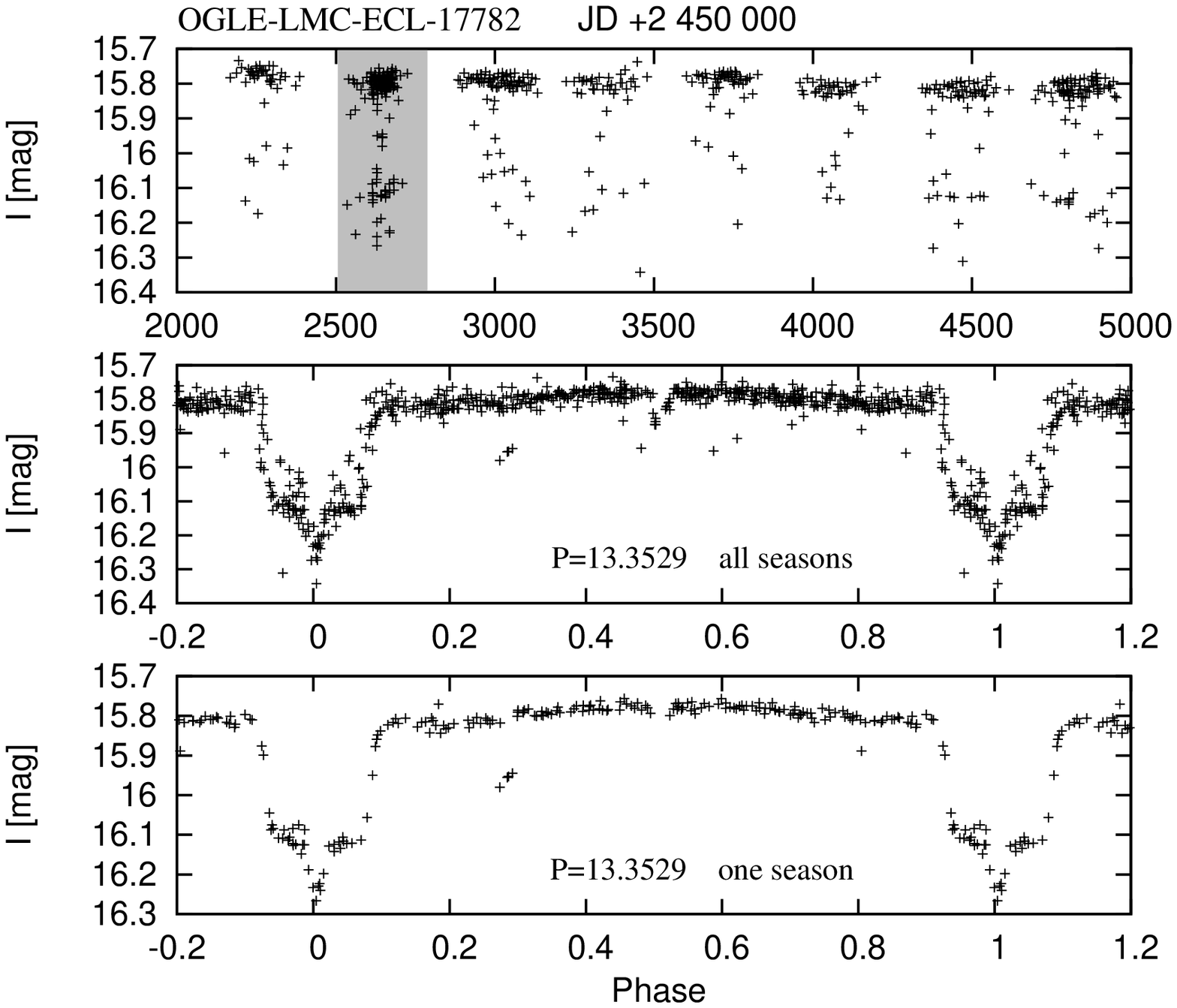}

\FigCap{Upper panel: light curve of OGLE-LMC-ECL-17782. Middle panel:
observations of OGLE-LMC-ECL-17782 folded with the period
$P=13.3529$~days reveal a strange looking light curve of an eclipsing
binary. The system configuration seems to be semidetached, but there are
some serious oddities to be explained -- see the text.  Bottom panel:
observations of OGLE-LMC-ECL-17782 from one OGLE-III observing season
(shaded region in the upper panel) again folded with the period $P$.
Note the well defined ``boxed'' primary minimum, suggesting that
variations of the shape occur from one season to another. There is also
startling, additional minimum of brightness visible near the phase 0.3,
but not visible during other observing seasons -- see the middle panel.}

\label{fig:13} 
\end{figure}

\begin{figure}[htb]
\includegraphics[scale=0.7]{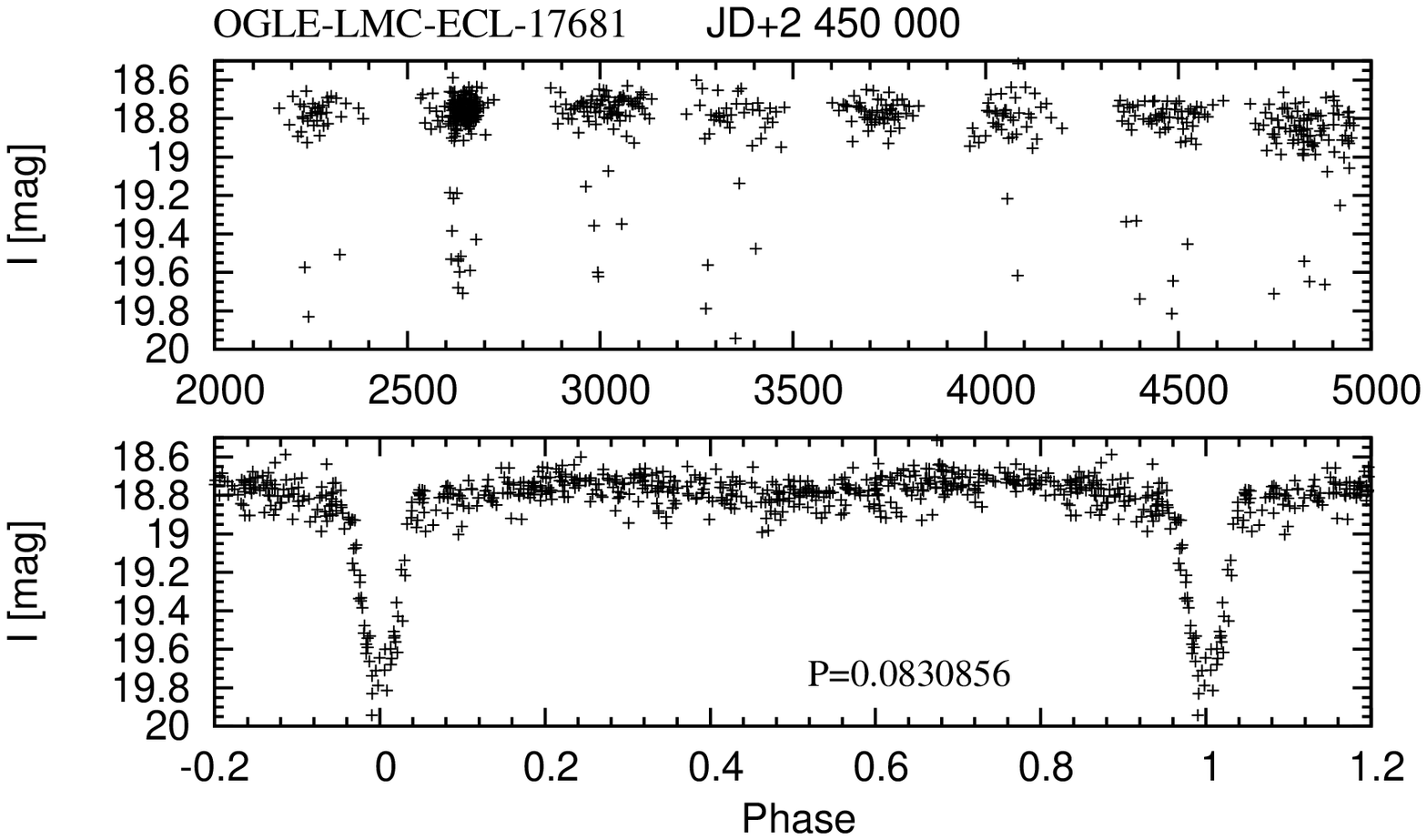}

\FigCap{Upper panel: light curve of OGLE-LMC-ECL-17681. Bottom panel:
light curve of OGLE-LMC-ECL-17681 folded with the period 0.0830856 days.
Note the deep primary minimum, the very shallow secondary minimum, small
reflection effect and no presence of eruptive or nova-like activity.}

\label{fig:14} 
\end{figure}

\begin{figure}[htb]
\includegraphics[scale=0.7]{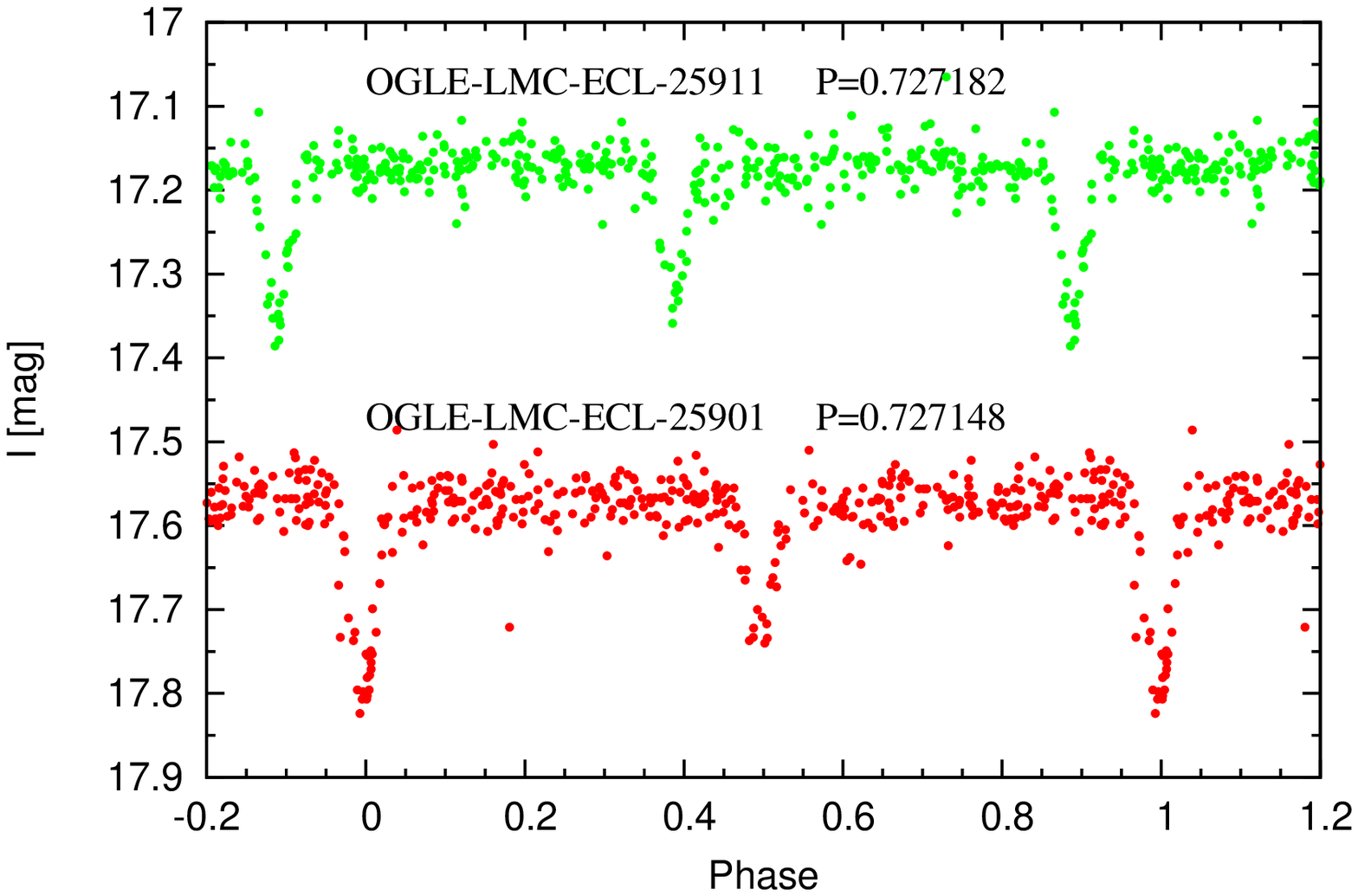}

\FigCap{Light curves of OGLE-LMC-ECL-25901 and OGLE-LMC-ECL-25911 folded
with their orbital periods and the same epoch of primary minimum. The
light curve of OGLE-LMC-ECL-25911 was shifted by $\sim-0.2$~mag for
clarity.}

\label{fig:15} 
\end{figure}

\section{Final Remarks}

Here we present the catalog of  eclipsing binary stars in the LMC based
on the OGLE-III photometric data, suitable for statistical analysis and
for individual case studies. This is the largest catalog of eclipsing
binaries published so far.

The catalog reveales a rich population of detached systems containing
early type components. Detached systems constitute most of the stars in
the catalog and their relative number is very similar to the relative
number of detached systems detected in the direction of the Cygnus-Lyra
region during the {\it Kepler} mission. 

The work on this catalog demonstrates that at some point of the
preparation a visual inspection of candidates was necessary. We feel,
however, that such procedure is close to the limit of practical sense.
During the verification of $79\;000$ candidate EBs we needed on average
13 seconds per star to make a reasonable evaluation (including the
period verification). For bright stars the procedure of visual
evaluation was quite fast, but for fainter stars, especially those close
to the limiting magnitude, visual inspection proved to be tedious. 

We think that the procedure of visual verification of candidates should
be replaced at this stage by some advanced algorithm which could
recognize the shape of the folded light curve, assign the proper period
and evaluate if this is an eclipsing binary. A final visual inspection
of ``the best'' sample (in our case: $\sim30\;000$ stars) would be much
more convenient. 

A promising step in that direction is the use of an artificial neural
network (ANN) like in Wyrzykowski \etal (2003, 2004). One of the
important factors during the evaluation of light curves by an ANN is the
use of a proper learning sample in order to teach an ANN how to
recognize which light curves corresponds to that of eclipsing binaries.
We consider that our catalog can be regarded as a comprehensive learning
sample for any ground based survey aimed at eclipsing binary detection. 
             
The OGLE-III catalog of eclipsing binary stars in the LMC is available
to the astronomical community from the OGLE Internet Archive:

\begin{center}
{\it http://ogle.astrouw.edu.pl}\\
{\it ftp://ftp.astrouw.edu.pl/ogle3/OIII-CVS/lmc/ecl/}
\end{center}

\Acknow{We would like to thank  Dr.~G. Pojma\'{n}ski for making
avalaible his code LC\_CLASS, Dr.~B. Pilecki for fruitful comments about
this work and Dr. Rory Smith for his English corrections to the text. 

The OGLE project has received funding from the European Research Council
under the European Community's Seventh Framework Programme
(FP7/2007-2013) / ERC grant agreement no. 246678 to AU. RP is
supported by the Foundation for Polish Science through the Start
Program. This publication was financed to DG by the GEMINI-CONICYT
Fund, allocated to the project N$^{o}$ 32080008.}

\end{document}